\documentclass[aps,12pt,pra,showpacs,preprint,a4paper]{revtex4}

\usepackage{amsmath}
\usepackage{graphicx}
\usepackage{subfig}
\usepackage{array}
\usepackage{amsfonts}
\usepackage{epsf}
\usepackage{epsfig}
\usepackage{amssymb}
\usepackage{bbm}
\usepackage{delarray}

\begin{document}

\title{A Laplace transform approach to find the exact solution of the $N$-dimensional Schr\"{o}dinger equation with Mie-type potentials and construction of Ladder operators }
\author{\small Tapas Das}
\email[E-mail: ]{tapasd20@gmail.com}\affiliation{Kodalia Prasanna Banga High
 School (H.S), South 24 Parganas, 700146, India}

\begin{abstract}
The second order $N$-dimensional Schr\"{o}dinger equation with
Mie-type potentials is reduced to a first order differential
equation by using the Laplace transformation. Exact bound state solutions are obtained using convolution or Faltungs theorem. The Ladder operators are also constructed for the Mie-type potentials in $N$- dimensions.
Lie algebra associated with these operators are studied and it is found that they satisfy
the commutation  relations for the SU(1,1) group.
\end{abstract}


\maketitle

\newpage

\section{Introduction}
The exact bound state solutions of the non-relativistic Schr\"{o}dinger equation with spherically symmetric potentials play an important role in atomic and molecular spectroscopy. Over the decades, theoretical physicists have shown a great deal of interest in solving multidimensional Schr\"{o}dinger equation for various spherically symmetric potentials [1-10].These higher dimension studies provide a general treatment of the problem in such a manner that one can obtain the required results in lower dimensions just dialing appropriate $N$. Many analytical as well as numerical technique have been developed by researcher to investigate multidimensional Schr\"{o}dinger equation for physically significant potentials[11-13]. \\
Mie-type potentials arise in the study of diatomic molecules.This type of potential is exactly solvable and studied by many authors for lower dimensional as well as higher dimensional Schr\"{o}dinger equation[14-17].The main goal of this paper is to use Laplace transform method and convolution or Faltungs theorem to solve the Mie-type potentials in $N$- dimensional Schr\"{o}dinger equation.The precious advantage of Laplace transformation is that it can convert the second order differential equation into first order. After achieving the first order differential equation the solution becomes very easy in transformed space. Finally we can reveal the actual solution in real space by using the inverse Laplace transformation method. Here the inverse transformation is done through the convolution of Faltungs theorem which is also a new aspect of this paper.        
More recent works on LT can be found in the ref [18-25].\\
The contents of this paper is as follows. Section II gives the brief introduction of the Laplace transform method, section III gives the energy eigenvalues and the energy eigenfunctions of the $N$ dimensional Schr\"{o}dinger equation with Mie-type potentials. In section IV Ladder operators for the considered potential as well as their Lie-algebra with Casimir operator have been studied. Finally section V serves the conclusion of the present work. 
\section{ Overview of Laplace transform method and convolution or Faltungs theorem}
The Laplace transform $\phi(s)$ or $\mathcal{L}$ of a function $f(t)$ is defined by [30,31]
\begin{eqnarray}
\phi(s)=\mathcal{L}\left\{{f(t)}\right\}=\int_{0}^{\infty}e^{-{st}}{f(t)}dt\,.
\end{eqnarray} 
If there is some constant $\sigma\in \Re$ such that ${\left|e^{-{\sigma}{t}}{f(t)}\right|\leq M}$
for sufficiently large $t$, the integral in Eq.(1) will exist for  Re $s>\sigma$ . The Laplace transform may fail to exist because of a sufficiently strong singularity in the function $f(t)$ as $t\rightarrow 0$ . In particular
\begin{eqnarray}
\mathcal{L}\left[\frac{t^{\alpha}}{\Gamma(\alpha+1)}\right]=\frac{1}{s^{\alpha+1}}\,,{\alpha}>-1\,.
\end{eqnarray}
The Laplace transform has the derivative properties 
\begin{eqnarray}
\mathcal{L}\left\{f^{(n)}(t)\right\}=s^n\mathcal{L}\left\{f(t)\right\}-\sum_{k=0}^{n-1}s^{n-1-k}{f^{(k)}(0)}\,,
\end{eqnarray}
\begin{eqnarray}
\mathcal{L}\left\{t^{n}f(t)\right\}=(-1)^{n}\phi^{(n)}(s)\,,
\end{eqnarray}
where the superscript$(n)$ denotes the 	$n$-th derivative with respect to $t$ for $f^{(n)}{(t)}$, and with respect to $s$ 
for $\phi^{(n)}{(s)}$.\\
The inverse transform is defined as $\mathcal{L}^{-1}\left\{\phi(s)\right\}=f(t)$. One of the most important properties of the Laplace transform is that given by the convolution of Faltungs theorem [30]. This theorem is a powerful method to find the inverse Laplace transform. According to this theorem if we have two transformed function $g(s)=\mathcal{L}\left\{G(t)\right\}$ and
$h(s)=\mathcal{L}\left\{H(t)\right\}$, then the product of these two is the Laplace transform of the convolution $(G*H)(t)$, where 
\begin{eqnarray}
(G*H)(t)=\int_{0}^t G(t-\tau)H(\tau)d\tau\,.
\end{eqnarray}
So Faltungs theorem yields
\begin{eqnarray}
\mathcal{L}(G*H)(t)=g(s)h(s)\,.
\end{eqnarray}
Hence 
\begin{eqnarray}
\mathcal{L}^{-1}\left\{g(s)h(s)\right\}=\int_{0}^t G(t-\tau)H(\tau)d\tau\,.
\end{eqnarray}
If we substitute $w=t-\tau$, then we find the important consequence $G*H=H*G$.

\section{Bound state spectrum}  
The Mie-type potentials[26] generally defined as
\begin{eqnarray}
V(r)=D_0\left[ \frac{a}{b-a}\left( \frac{r_0}{r}\right)^b-\frac{b}{b-a}\left( \frac{r_0}{r}\right)^a\right] \,,
\end{eqnarray} 
where $D_0$ is the interaction energy between two atoms in a molecular system at equilibrium distance $ (r=r_0) $. Parameters $a=2,b=1$ give standard Morse or Kratzer-Fues potential of the form [27]
\begin{eqnarray}
V(r)=-D_0\left(\frac{2r_0}{r}-\frac{r_0^2}{r^2}\right)\,.
\end{eqnarray} 
Moreover, the standard Kratzer potential is modified by adding a term to get the modified Kratzer-type potential[27]
\begin{eqnarray}
V(r)=-D_0\left(\frac{r-r_0}{r}\right)^2 \,.
\end{eqnarray}
It is customary to take a general form of Mie-type potentials as
\begin{eqnarray}
V(r)=\frac{A}{r^2}+\frac{B}{r}+C\,.
\end{eqnarray}
This form is more flexible because with $A=-D_0r_0^2, B=2D_0r_0 $ and $C=-D_0$, we have the modified Kartzer potential and similarly  for $A=D_0r_0^2, B=-2D_0r_0 $ and $C=0$ we have Kartzer-Fues potential. \\
The time independent Schr\"{o}dinger equation for a particle of mass
$M$ in $N$-dimensional space has the form [28]
\begin{eqnarray}
-\frac{\hbar^2}{2M}\nabla^2_{N}\psi+V\psi=E\psi\,,
\end{eqnarray}
where $\nabla^2_{N}$ is the Laplacian operator in the polar
coordinates $(r, \theta_1, \theta_2, \ldots,
\theta_{N-2},\varphi)$ of $R^{N}$. Here $r$ is the hyperradius and
$\theta_1, \theta_2, \ldots, \theta_{N-2},\varphi$ are the
hyperangles. The form of $\nabla^2_{N}$ is given by
\begin{eqnarray}
\nabla^2_{N}=r^{1-N}\frac{\partial}{\partial
r}\left(r^{N-1}\frac{\partial}{\partial
r}\right)+\frac{\Lambda^2_{N}(\Omega)}{r^2}\,,
\end{eqnarray}
where $\Lambda^2_{N}(\Omega)$ is the hyperangular momentum
operator [28] given by
\begin{equation*}
\Lambda_N^2 = - \sum_{\stackrel{i,j=1}{i>j}}^N \Lambda_{ij}^2 \;, \;\; \Lambda_{ij} = x_i \frac{\partial}{\partial x_j} - x_j \frac{\partial}{\partial x_i}\,,
\end{equation*}
for all Cartesian components $x_i$ of the $N$-dimensional vector $(x_1,x_2, \dots, x_N)$.
Radial part of Eq.(12) is extracted by using separation variable method. The separation constant in this purpose is taken 
as $\beta_{\ell N}(N>1)=\ell(\ell+N-2)$  with $\ell=0, 1,2, \ldots$, [29]. In this way the $N$-dimensional hyperradial or in short the ``radial" Schr\"{o}dinger equation becomes
\begin{eqnarray}
\left[\frac{d^2}{dr^2}+\frac{N-1}{r}\frac{d}{dr}-\frac{\ell(\ell+N-2)}{r^2}+\frac{2M}{\hbar^2}[E-V(r)]\right]R(r)=0\,,
\end{eqnarray}
where $E$ is the energy eigenvalue and $\ell$ is the orbital
angular momentum quantum number.\\
Inserting the Mie-type potentials given by Eq.(11) into the Eq.(14) and taking the following abbreviations 
\begin{eqnarray}
\nu(\nu+1)=\ell(\ell+N-2)+\frac{2MA}{\hbar^2}\,\,; \frac{2M}{\hbar^2}(E-C)=-\epsilon^2\,\,; \frac{2MB}{\hbar^2}=-\beta 
\end{eqnarray}
with $R^{''}(r)=\frac{d^2R}{dr^2}\,\,;R^{'}(r)=\frac{dR}{dr}$
we have 
\begin{eqnarray}
R^{''}(r)+\frac{N-1}{r}R^{'}(r)-\frac{\nu(\nu+1)}{r^2}R(r)-\epsilon^2R(r)+\frac{\beta}{r}R(r)=0\,.
\end{eqnarray}
Let us consider the bound state solution like $R(r)=r^{-k}f(r)$ with  $k>0$. Here the term $r^{-k}$ ensures that for
$r\rightarrow\infty$, $R(r)\rightarrow 0$ and $f(r)$ is expected to behave like $f(r)\rightarrow 0$ as $r\rightarrow 0$. Now inserting the above assumed solution in Eq.(16) it is easy to achieve
\begin{eqnarray}
rf^{''}(r)+(N-2k-1)f^{'}(r)+\left\{\frac{k(k+1)-k(N-1)-\nu(\nu+1)}{r}-\epsilon^2 r+\beta\right\}f(r)=0 \,,
\end{eqnarray}
were double prime over $f(r)$ denotes the second order derivative of $f(r)$ with respect to $r$ and similarly the single prime over the same denotes the first order derivative. To get a Laplace transform of the above equation we need a parametric restriction
\begin{eqnarray}
k(k+1)-k(N-1)-\nu(\nu+1)=0\,,
\end{eqnarray}
which has a solution $k_+=k_{\ell N}$. So here we have
\begin{eqnarray}
rf^{''}(r)+(N-2k_{\ell N}-1)f^{'}(r)+\left\{-\epsilon^2 r+\beta\right\}f(r)=0 \,.
\end{eqnarray}
Hence identifying $t$ as $r$ in Eq.(1) i.e $\phi(s)=\mathcal{L}\left\{{f(r)}\right\}$ and using Eq.(3-4) in the same manner, it is easy to get 
\begin{eqnarray}
(s^2-\epsilon^2)\frac{d\phi(s)}{ds}+\left\{s(2k_{\ell N}-N+3)-\beta\right\}\phi(s)=0\,.
\end{eqnarray}  
This is a linear first order homogeneous differential equation, which has a simple solution [24] of the form
\begin{eqnarray}
\phi(s)=D (s+\epsilon)^{-(2k_{\ell N}-N+3)}\left(1-\frac{2\epsilon}{s+\epsilon}\right)^{\frac{\beta-(2k_{\ell N}-N+3)\epsilon}{2\epsilon}}\,,
\end{eqnarray}
were $D$ is a constant.The expression $\left(1-\frac{2\epsilon}{s+\epsilon}\right)^{\frac{\beta-(2k_{\ell N}-N+3)\epsilon}{2\epsilon}}$ is a multivalued function. The wave functions must be single valued, so we must take
\begin{eqnarray}
\frac{\beta-(2k_{\ell N}-N+3)\epsilon}{2\epsilon}=n\,, n=0,1,2,3, \ldots
\end{eqnarray}
In this manner from Eq.(21),we have 
\begin{eqnarray}
\phi(s)=D(s+\epsilon)^{-a}(s-\epsilon)^{-b}=Dg(s)h(s)\,,
\end{eqnarray}
where $a=2k_{\ell N}-N+3+n$ and   $b=-n$. In order to find $f(r)=\mathcal{L}^{-1}\left\{\phi(s)\right\}$, we find [31]
\begin{eqnarray}
\mathcal{L}^{-1}\left\{(s+\epsilon)^{-a}\right\}=G(r)=\frac{r^{a-1}e^{-\epsilon r}}{\Gamma (a)}\,,\nonumber\\
\mathcal{L}^{-1}\left\{(s-\epsilon)^{-b}\right\}=H(r)=\frac{r^{b-1}e^{\epsilon r}}{\Gamma (b)}\,.
\end{eqnarray}
Now using Eq.(7), we have
\begin{eqnarray}
f(r)=\mathcal{L}^{-1}\left\{\phi(s)\right\}=D(G*H)(r)=D \int_{0}^r G(r-\tau)H(\tau)d\tau\nonumber\\
=\frac{De^{-\epsilon r}}{\Gamma(a)\Gamma(b)}\int_{0}^r (r-\tau)^{a-1}\tau^{b-1} e^{2\epsilon \tau}d\tau\,.
\end{eqnarray}
The integration can be found in [32], which gives
\begin{eqnarray}
\int_{0}^r (r-\tau)^{a-1}\tau^{b-1} e^{2\epsilon \tau}d\tau=B(a,b)r^{a+b-1}\,_{1}F_{1}(b, a+b, 2\epsilon r)\,,
\end{eqnarray}
where $\,_{1}F_{1}$ is confluent hypergeometric functions [30]. Now using the Beta function\\ $B(a,b)=\frac{\Gamma (a)\Gamma(b)}{\Gamma(a+b)}$, the final form of $f(r)$ can be written from Eq.(25)
\begin{eqnarray}
f(r)=\frac{D}{\Gamma(a+b)}e^{-\epsilon r}r^{a+b-1}\,_{1}F_{1}(b, a+b, 2\epsilon r)\,.
\end{eqnarray} 
So the radial wave function can be given as
\begin{eqnarray}
R_{n\ell N}(r)=r^{-k_{\ell N}}f(r)=\zeta_{n\ell N}r^{(k_{\ell N}+2-N)}e^{-\epsilon r}\,_{1}F_{1}(-n,2k_{\ell N}+3-N,2\epsilon r)\,, 
\end{eqnarray}
were $\zeta_{n\ell N}=\frac{D}{\Gamma(a+b)}$ is the normalization constant can be evaluated from the condition
\begin{eqnarray}
\int_{0}^\infty |R_{n\ell N}(r)|^2 r^{N-1}dr=1\,.
\end{eqnarray}
To evaluate the integration here we have some useful formulas [30]
\begin{eqnarray*}
\,_{1}F_{1}(-q,\alpha+1,\gamma)=\frac{q!\alpha!}{(q+\alpha)!}L_{q}^{\alpha}(\gamma)\,,
\end{eqnarray*}
and
\begin{eqnarray*}
\int_{0}^\infty x^{w+1}e^{-x}\left\{{L_{h}^w}(x)\right\}^2dx=\frac{(w+h)!}{h!}(2h+w+1)\,.
\end{eqnarray*}
Hence the normalization constant becomes
\begin{eqnarray}
\zeta_{n\ell N}=(2\epsilon)^{k_{\ell N}+2-\frac{N}{2}}{\frac{1}{(2k_{\ell N}+2-N)!}}\sqrt{\frac{(2k_{\ell N}+2-N+n)!}{n!(2k_{\ell N}+2n+3-N)!}}\,.
\end{eqnarray}
Finally we write the energy eigenvalues from Eq.(22) and Eq.(15) as
\begin{eqnarray}
E_{n\ell N}=C-\frac{M}{2\hbar^2}\left(\frac{B}{n+k_{\ell N}+\frac{3-N}{2}}\right)^2\,,
\end{eqnarray}
with the eigenfunctions
\begin{eqnarray}
R_{n\ell N}(r)=(2\epsilon)^{k_{\ell N}+2-\frac{N}{2}}{\frac{1}{(2k_{\ell N}+2-N)!}}\sqrt{\frac{(2k_{\ell N}+2-N+n)!}{n!(2k_{\ell N}+2n+3-N)!}}\nonumber\\
\times r^{(k_{\ell N}+2-N)}e^{-\epsilon r}\,_{1}F_{1}(-n,2k_{\ell N}+3-N,2\epsilon r)\,. 
\end{eqnarray}
The complete orthonormalized	 energy eigenfunctions of the
$N$-dimensional Schr\"{o}dinger equation with Mie-type potentials
can be given by
\begin{eqnarray}
\psi(r, \theta_{1}, \theta_{2}, \ldots, \theta_{N-2},
\phi)=\sum_{n, \ell,
m}\zeta_{n\ell N}R_{n\ell N}(r)Y_{\ell}^{m}(\theta_{1}, \theta_{2},
\ldots, \theta_{N-2}, \phi)\,,
\end{eqnarray}
where $Y_{\ell}^{m}(\theta_{1}, \theta_{2}, \ldots, \theta_{N-2},
\phi) \equiv Y_{\ell}^{m}(\Omega)$ are the hyperspherical
harmonics of degree $\ell$ on the $S^{N-1}$ sphere. These
harmonics are the root of the equation
\begin{eqnarray}
\Lambda^{2}_{N}(\Omega)Y_{\ell}^{m}(\Omega)+\ell(\ell+N-2)Y_{\ell}^{m}(\Omega)=0\,,
\end{eqnarray}
which is the separated part of Eq.(12). 

\section{Construction of Ladder Operators for Mie-type Potentials in $N$-Dimensions}
In this part of the paper, Ladder operators for Mie-type potentials have been constructed from the eigenfunctions that obtained in Eq.(26).Using the formula given just after Eq.(29), the Eq.(26) can be written as
\begin{eqnarray}
R_{n\ell N}(y)=\eta_{n\ell N}y^{k_{\ell N}+2-N}e^{-\frac{y}{2}}L_{n}^{2k_{\ell N}+2-N}(y)\,, 
\end{eqnarray}
where $\eta_{n\ell N}=\zeta_{n\ell N}\left(\frac{1}{2\epsilon}\right)^{k_{\ell N}+2-N}\frac{n!(2k_{\ell N}+2-N)!}{(2k_{\ell N}+n+2-N)!}$ and $y=2\epsilon r$.
Our goal is to find the differential Operators $\hat{L}_{\pm}$
satisfying the property
\begin{eqnarray}
\hat{L}_{\pm}R_{n\ell N}(y)=\lambda_{\pm}R_{n\pm{1},\ell N}(y) \,. 
\end{eqnarray}  
In other word, we wish to find the operators of the form
\begin{eqnarray}
\hat{L}_{\pm}=f_\pm(y)\frac{d}{dy}+g_\pm (y) \,.
\end{eqnarray}
It is easy to achieve the following equation
\begin{eqnarray}
y\frac{d}{dy}R_{n\ell N}=(k_{\ell N}+2-N)R_{n\ell N}-\frac{y}{2}R_{n\ell N}+\eta_{n\ell N}y^{k_{\ell N}+2-N}e^{-\frac{y}{2}}y\frac{d}{dy}L_{n}^{2k_{\ell N}+2-N}(y)\,.
\end{eqnarray}
Applying the recurrence relation associated with Laguerre polynomials [30]
\begin{eqnarray}
x\frac{d}{dx}L_{n}^\alpha (x)=n L_{n}^\alpha (x)-(n+\alpha)L_{n-1}^\alpha (x) \,,
\end{eqnarray}
we obtain
\begin{eqnarray}
\left(-y\frac{d}{dy}+k_{\ell N}+n+2-N-\frac{y}{2}\right)R_{n\ell N}=\frac{\eta_{n\ell N}}{\eta_{n-1,\ell N}}(n+k_{\ell N}+2-N)R_{n-1,\ell N}\,.
\end{eqnarray}
So we have the annihilation operator
\begin{eqnarray}
\hat{L}_{-}=-y\frac{d}{dy}-\frac{y}{2}+k_{\ell N}+n+2-N\,,
\end{eqnarray}
with eigenvalues
\begin{eqnarray}
\lambda_-=\sqrt{\frac{n(n+2k_{\ell N}+2-N)(2n+1+2k_{\ell N}-N)}{(2n+2k_{\ell N}+3-N)}}\,.
\end{eqnarray}
Similarly following the recurrence relation [30]
\begin{eqnarray}
x\frac{d}{dx}L_{n}^\alpha (x)=(n+1) L_{n+1}^\alpha (x)-(n+\alpha+1-x)L_{n}^\alpha (x) \,,
\end{eqnarray}
we have the creation operator
\begin{eqnarray}
\hat{L}_{+}=y\frac{d}{dy}-\frac{y}{2}+n+k_{\ell N}+1\,,
\end{eqnarray}
with eigenvalues
\begin{eqnarray}
\lambda_+=\sqrt{\frac{(n+1)(n+2k_{\ell N}+3-N)(2n+2k_{\ell N}+5-N)}{(2n+2k_{\ell N}+3-N)}}\,.
\end{eqnarray}
Here it is interesting to study the Lie algebra associated with the operators $\hat{L}_{\pm}$. Using the set of Eq.(41,42) and 
Eq.(44,45) we can compute commutator $[\hat{L}_{-},\hat{L}_{+}]$ as
\begin{eqnarray}
[\hat{L}_{-},\hat{L}_{+}]R_{n\ell N}(y)=\lambda_0 R_{n\ell N}(y)\,,
\end{eqnarray}
with the eigenvalues 
\begin{eqnarray}
\lambda_0=2n+2k_{\ell N}-N+3\,.
\end{eqnarray}
This makes possible to construct the operator
\begin{eqnarray}
\hat{L}_{0}=\hat{n}+k_{\ell N}+\frac{3-N}{2}\,.
\end{eqnarray}
Now it is easy to compute the commutator relations among the three operators $\hat{L}_{\pm}$ and $\hat{L}_{0}$. They satisfy the following Lie algebra
\begin{eqnarray}
[\hat{L}_{-},\hat{L}_{+}]=2\hat{L}_{0}  \,\,;
[\hat{L}_{0},\hat{L}_{+}]=\hat{L}_{+} \,\,;
[\hat{L}_{-},\hat{L}_{0}]=\hat{L}_{-}\,,
\end{eqnarray}
which corresponds to the commutator relations of the SU(1,1) algebra. 
We can generate the following commutator brackets using the relations given in Eq.(49) 
\begin{eqnarray}
[\hat{L}_{0},\hat{L}_{a}]=\hat{L}_{s} \,\,; [\hat{L}_{0},\hat{L}_{s}]=\hat{L}_{a}\,,
\end{eqnarray}
where $\hat{L}_{a}=\hat{L}_{+}+\hat{L}_{-}$  and  $\hat{L}_{s}=\hat{L}_{+}-\hat{L}_{-}$. \\
Finally, the Casimir operator [33] of the group can also be expressed as
\begin{eqnarray}
\tilde{C}=\hat{L}_{0}(\hat{L}_{0}-1)-\hat{L}_{+}\hat{L}_{-}=\hat{L}_{0}(\hat{L}_{0}+1)-\hat{L}_{-}\hat{L}_{+}\,,
\end{eqnarray}
with the eigenvalue equation 
\begin{eqnarray}
\tilde{C}R_{n\ell N}=J(J-1)R_{n\ell N}\,;
\end{eqnarray} 
where $J=k_{\ell N}+\frac{3-N}{2}$.
\section{Conclusions}
Some aspects of $N$- dimensional hyperradial
Schr\"{o}dinger equation for Mie-type potentials have been investigated by Laplace transform
approach. It is found that the energy eigenfunctions and the
energy eigenvalues depend on the dimensionality of the problem. This paper is flexible in the sense that we can obtain the results of some special cases. Coulomb potential, with $A=C=0$ and for arbitrary $N$, the results agree with the ref [24]. We can reach all the results for Mie-type potentials, that obtained in the work listed in reference [20] just finding $k_{\ell 3}	(>0)$ from Eq.(18) in association with Eq.(15) for the special case $N=3$ . The bound state spectrum of Kartzer-Fues potential can be achieved by setting $C=0$ for ordinary three dimensional system as well as for arbitrary dimensions.
Moreover, using the recurrence relations associated with the Laguerre polynomials,creation and annihilation operators have been constructed for the Mie-type potentials in $N$- dimensions and they are found to agree with the Lie algebra of SU(1,1) group. Also Casimir operator of the group has been studied at the end.

\end{document}